\documentclass[prb,twocolumn,showpacs,floatfix]{revtex4}
\usepackage{latexsym}
\usepackage{graphicx}

\usepackage{subfigure}    
\usepackage{epsfig}
\begin{document}
\title{Photovoltaic effect for narrow-gap Mott insulators}
\author{Efstratios Manousakis}
\affiliation{Department of Physics,
Florida State University,
Tallahassee, FL 32306-4350, USA\\
and Department of  Physics, University of Athens,
Panepistimioupolis, Zografos, 157 84 Athens, Greece
}
\date{\today}
\begin{abstract}
We discuss the photovoltaic effect at a p-n heterojunction,
in which the illuminated side is a doped Mott insulator, using
the simplest description of a Mott insulator within the 
Hubbard model. We find that the internal quantum efficiency
of such a device, if we choose an appropriate {\it narrow-gap} 
Mott insulator, can be significantly enhanced due to impact 
ionization caused by the photoexcited ``hot'' electron-hole pairs. 
Namely, the photoexcited electron and/or hole
can convert its excess energy beyond the Mott-Hubbard
gap to additional electrical energy by creating multiple
electron-hole pairs in a time scale which can be shorter than  the
time characterizing other relaxation processes.
\end{abstract}
\pacs{71.10.Fd,71.27.+a,72.40.+w,73.50.Pz,78.56.-a,84.60.Jt,88.40.H-,88.40.J-}
\maketitle
%\section{The model}
 %Give and introduction to the problem where you include the literature
%on ``standard semiconducting'' materials. Discuss the issue of
%internal or quantum efficiency.
\section {Introduction} 
Solar cells based on  conventional single-gap 
band-semiconductors  have low internal quantum efficiency for conversion
of solar into electrical energy \cite{Shockley}. 
The main reason is that
the  photon energy absorbed by a single electron-hole pair with energy
above the semiconductor band
gap is lost in heat through electron-phonon scattering and phonon emission
through which electrons and holes relax to their band edges.
The solar energy spectrum contains photons with energies in a rather broad 
energy range compared to any particular semiconductor energy gap. 
If instead semiconductors with a relatively large band-gap 
are chosen in order to avoid the above mentioned problem, this cuts off most
of the energy spectrum of the solar radiation.
Using a stack of cascaded multiple p-n junctions with various
band gaps is an expensive approach to build efficient solar
cells.

A variety of other ideas have been suggested in order to 
increase solar cell efficiency \cite{nozik,Ross,Crabtree,Nori}. 
%One approach is to extract the carriers before
%they ``cool'' by phonon emission and this produces an enhanced
%photo-voltage. 
One particular approach which increases the photo-current
is one in which the energetic carriers produce additional
electron-hole pairs through impact ionization \cite{impact0,impact1,impact2}
 at a much faster
rate than the rate characterizing the relaxation process through
phonon emission. 
In order to achieve this, it has been suggested that \cite{nozik1} 
using confining geometries, such as quantum wells,
quantum wires, quantum dots, superlattices and nanostructures,
the relaxational time scales can be significantly 
affected \cite{Benisty,Bockelmann,Benisty2}
and this allows the possibility for impact ionization.
Ross and Nozik \cite{Ross} have argued that an efficiency of
approximately $66\%$, much larger than Shockley-Queisser efficiency
limit \cite{Shockley} for photovoltaics, can be theoretically 
achieved using an unusual 
inversion, in which 
the chemical potential of the excited electronic band is below that of the 
ground band. Other ideas to increase solar-cell efficiency
by utilizing the possibility of carrier multiplication 
\cite{Werner,Brendel,Spirkl} via
impact ionization by removing and isetropically cooling the carriers through 
special membranes were proposed by W\"urfel\cite{Wurfel}.

In this paper we discuss the photovoltaic effect for  
materials which  are based on doped Mott insulators, as opposed to
the familiar semiconductors which are band-insulators. 
A variety of transition metal oxides that are predicted to 
be conductors by band theory (because they have an odd number of electrons 
per unit cell) are, in fact, insulators. Mott and Peierls \cite{Mott} predicted 
that this anomaly can be explained by including interactions between 
electrons. Today, this can be described within the simple Hubbard
model, as arising from the strong on-site Coulomb interaction. In this case
the half-filled band leads to an insulator; the added holes or
added extra electrons beyond half filling lead to conduction.
In this case also we can imagine that we can form a p-n junction separating
a region of hole-doped from a region of electron-doped 
Mott insulator. 

There are recent reports  \cite{HLiu,Sun,Qiu,Luo} 
where the photovoltaic
effect is observed in a p-n heterojunction of a doped magnanite (a doped 
Mott-like  insulator) and doped
strontium titanate (a doped band insulator). 
In addition, there are reports of the photoelectric
effect observed on heterojunctions of doped magnanites 
and doped silicon \cite{HLu,KZhao} and on a heterostructure \cite{Muraoka} 
of $YBa_2Cu_3O_{7-x}$ on Nb doped  $SrTiO_3$. 
Here, we demonstrate  that in the case of 
{\it narrow-gap} Mott insulators the photovoltaic 
effect can lead to solar
cells of high quantum efficiency, where a single solar photon
can produce multiple electron-hole pairs.

\section{Model and concepts}

\begin{figure}[htp]
\includegraphics[width=2.5 in]{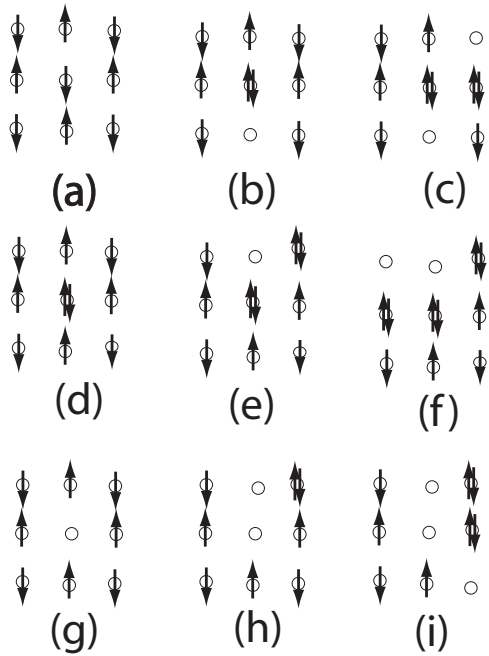}
\caption{Representative states from the $S^n_m$ subspaces.}
\label{fig1}
\end{figure}

In the simplest model of a Mott insulator, 
a single conduction band is formed from one
localized orbital (which is, in general, a linear combination
of atomic orbitals) per unit cell (or site). 
In the case of half-filling, i.e., when
one electron per site occupies the band made from this orbital,
the material is a Mott insulator. In Fig.~\ref{fig1}(a) 
the half-filled state is shown where the motion of an electron  from one 
site to the nearest-neighbor site 
leads to an empty site (or hole) and a doubly-occupied
site (DO) (Fig.~\ref{fig1}(b)); 
this state, because of the DO, is characterized by a high energy cost, 
which is denoted
by the so-called Hubbard on-site Coulomb repulsion energy $U$, and this
effectively prohibits electron motion through the band. 
In order to have a concrete framework to facilitate the 
discussion we consider the so-called 
Hubbard model.
\begin{eqnarray}
\hat H &=& \hat H_t + \hat H_U, \\
H_t &=& -t \sum_{<ij>} (c^{\dagger}_{j\sigma} c_{i\sigma} + h.c), \\
\hat H_U &=& U \sum_i \hat n_{i\uparrow} n_{i\downarrow},
\end{eqnarray}
where $n_{i\sigma} = c^{\dagger}_{i\sigma} c_{i\sigma}$ and 
$c^{\dagger}_{i\sigma}$ creates an electron of spin $\sigma$ at site $i$.

%\begin{figure}[htp]
%\includegraphics[width=2.0 in]{Fig2.eps}
%\caption{Schematic illustration of the density of states
%of the Hubbard model for electron (DO sites) and holes.
%The energy gap is denoted by a parameter $U^*$.}
%\label{fig2}
%\end{figure}

%\begin{figure}[htp]
%\includegraphics[width=2.5 in]{densityofstates.eps}
%\caption{Schematic illustration of the density of states
%of the Hubbard model for single-hole excitations (left) and
%single DO excitations.}
%\label{fig2.b}
%\end{figure}

First, let us consider the limit of $U \gg t$,
because in this case there is a well-defined concept for the origin of the
Mott-insulator gap and this simplifies the conceptual part of the 
discussion. 
If we assert that the $\hat H_U$ term is the unperturbed part and
we treat $\hat H_t$ as a perturbation, the eigenvalues
of the unperturbed part are equally spaced by an energy $U$.
For the case of half-filling, i.e., in an $N$-site system with $N$ electrons,  
each energy level corresponds to a massive number of states
forming subspaces $S^{n}_0$ characterized by the same eigenvalue $E_n = nU$,
where $n$ is the number of pairs of  DO-holes (DO-H).
Figs.~\ref{fig1}(a),(b) and (c) illustrate examples of 
states $n=0,1$ and 2 pairs of DO-H. 
The subspaces $S^{n}_0$ are spanned
by the states obtained by all allowed spin rearrangements. 
In a Mott insulator, the Coulomb energy cost for
two electrons to occupy the same orbital leads to an
insulating state at half-filling.
Replacing some atoms in the lattice with different atoms,
which act as donors or acceptors of additional electrons 
beyond half-filling thereby
creating DO sites (n-type doping) or holes (p-type doping), 
leads to a conducting state with an
effective carrier density given, respectively, by the density of DO 
or hole sites.
In a doped system, i.e., an $N$-site system with $N+m$ or $N-m$ electrons,
the subspaces $S^{n}_m$ or $S^{n}_{-m}$ are spanned by
$m$ DO or holes and $n$ DO-H pairs. Examples for $S^n_1$ and $S^n_{-1}$ 
(for $n=0,1,2$) are
 illustrated in Figs.~\ref{fig1}(d), (e), and (f), and 
in Figs.~\ref{fig1}(g) (h), and (i). 
%The density of states for an excited electron (DO site) and hole is
%shown in Fig.~\ref{fig2}.
For the rest of the paper, we will denote by $U^*$ the
renormalized gap, namely, the energy required to promote a particle
and to create a DO-H pair starting from the half-filled lattice
and by $W$ the bandwidth of the quasiparticle or quasihole dispersion.

%If we restrict ourselves in one specific subspace
%$S^{n}_m$ (or $S^n_{-m}$), with $N+m$ (or $N-m$) electrons and, in 
%addition, $n$ DO/H pairs, one can carry out up 
%to second order degenerate perturbation theory to derive the well-known
%$t-J$ model (with a three-site term which is usually
%neglected). 
In the half-filled case and at low enough temperature, 
the model exhibits antiferromagnetic order \cite{RMP}, which, however, 
will not play a significant role in our discussion, because the
Mott-insulator phase can exist at temperatures higher than the
N\'eel ordering temperature.

\begin{figure}[htp]
\includegraphics[width=2.5 in]{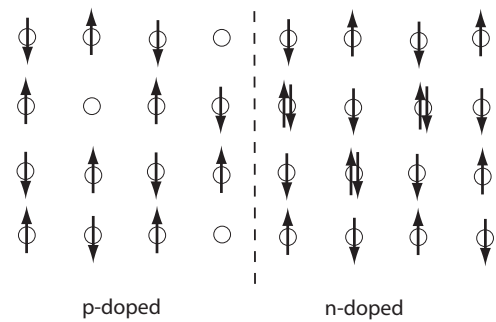}
\caption{An interface between a p- and an n-doped Mott 
insulator before carrier diffusion
across the interface takes place.
Some of the DO sites can be relieved from 
the energy cost due to Coulomb repulsion via diffusion of the
extra electron to the p-doped region to fill a hole. 
This process stops when  the electric field, which gradually develops 
across the interface due to this charge redistribution, becomes strong 
enough to oppose further carrier diffusion. The idea discussed here
is also applicable when the illuminated side is a doped Mott-insulator
and the other side a doped band-insulator.}
\label{fig2}
\end{figure}

The quasiparticle (hole or DO) spectral function and dispersion relation, as
well as the spectral weights, have been computed using 
the $t-J$ model \cite{Liu}
(which is the strong coupling limit ($U \gg t$) of the Hubbard model
with $J=4t^2/U$)
within the so-called non-crossing approximation for 2D.
In this paper we have carried out the same calculation for
3D and the conclusions are qualitatively similar to the 2D case.
There is a well-defined quasiparticle which corresponds to
a low energy peak in the spectral function, and we will use the
operator $\alpha^{\dagger}_{\bf k}$ which, by acting on the undoped 
state $| 0 \rangle$, creates this 
quasiparticle with energy dispersion $E_{\bf k}$ and  residue $Z_{\bf k}$.
%The density of states of the Hubbard model for a quasihole or a quasiparticle
%(a state with a single mobile DO site) is shown schematically in
%Fig.~\ref{fig2.b}. 
In this limit, while the bandwidth of the quasiparticle
dispersion $E_{\bf k}$ is of the order of $J$, the 
width of the density of states is of the order of $2 t d$ (where $d$ is the
dimensionality).

%\section{ Description of the mechanism}
Now we wish to consider a p-n heterojunction as illustrated in Fig.~\ref{fig2}
based on an interface between two  narrow-gap 
Mott insulators, one of which is p-doped (left side of the interface) and 
the other which is n-doped (right side). 
Because of the energy gain that is achieved when a hole
of the p-doped material is filled by an electron from a DO
site on the n-doped part, diffusion of electrons from the
n-doped side  will occur and this leads to a positively charged
layer in the n-doped side adjacent to the interface and, vice versa,
a negatively-charged layer in the p-doped side. 
The potential energy difference acts as 
a unidirectional collector of the DO and holes created upon the
incident light.

\begin{figure}[htp]
\includegraphics[width=2.5 in]{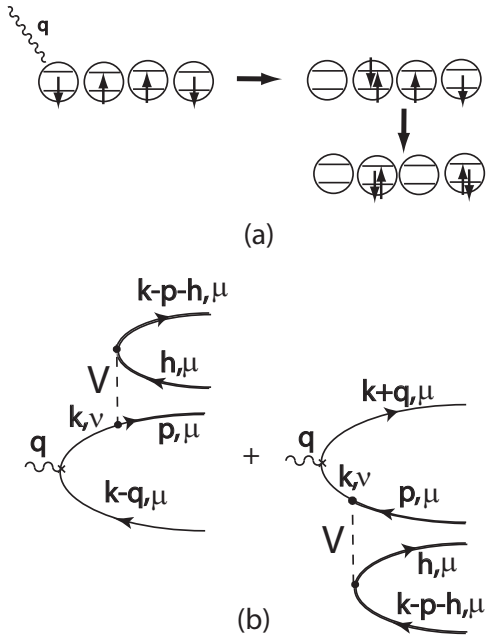}
\caption{ See the text for explanation.}
\label{fig3}
\end{figure}

\section{ Narrow-gap narrow-band case} 
\label{narrowband}

Let us, first, consider a narrow-band narrow-gap Mott insulator. 
An incident photon of energy several times the Mott-gap $U^*$ has enough energy
so that several electron-hole pairs could be created. 
In practice, since the range of the solar spectrum is from
$0.5-3.5$ eV, we can consider a Mott-insulator with a gap of
the order of $0.5-1.0$ eV and a bandwidth in the same range.
We need a narrow-gap Mott insulator in order
to increase the efficiency, but on the other hand the gap should be
large enough to avoid decay of the excited quasiparticles (by phonon 
emission, spin-wave emission or through other possible 
excitations).
%The fundamental process that couples the incident photon with more
%than one electrons at the same time is the fact that
%the initially created spatially separated double-occupancy site
%and the site with the hole are charged atoms and they
%couple with their neighboring atoms with a large interaction
%coupling. This  is not the case in standard semiconductor.

If the energy $\hbar \omega_q$ of the solar photon is between 
$U^*<\hbar \omega_q<U^*+W$,
the incident photon can create a single electron-hole pair by promoting
an electron to the first Mott-Hubbard band. 
Fig.~\ref{fig3}(a) indicates a two-step process in which the incident photon 
can create a state with two pairs of DO-H, if its energy is higher 
than $2U^*$. In Fig.~\ref{fig3}(a), within 
each atom we depict two atomic levels, one which forms the Mott-Hubbard
band, and a higher level which forms a  high-energy band.
First, the incident high-energy photon is absorbed by an electron
which hops to the high-energy band as
indicated in Fig.~\ref{fig3}(a). 
Subsequently, the charged atom, which carries the extra electron, transfers
part of the excess energy via the Coulomb interaction 
to an electron of a different site
which causes it to hop to a n.n. atom, while itself falls to the lowest level
already occupied by one electron; this leads to the creation 
of two DO-H pairs. 
Fig.~\ref{fig3}(b) illustrates the same process in momentum space.
If the energy of the photon
is higher than $2U^*$, the electron (or the hole) can 
be temporarily photo-excited  to a  high-energy band which is different 
than the Mott band. The electron, for example, can be
photo-excited to the state 
$|{\bf k} ~\nu \rangle$ of energy $E$ denoted by the label 
${\bf k},\nu$ in the first term
of Fig.~\ref{fig3}(b). This electron
can decay quickly into two quasiparticles plus one quasihole
state of the Mott band, provided that the energy of the 
initial and final state match.  This is caused by the matrix element 
$V\equiv \langle {\bf k} ~\nu, {\bf h} ~\mu | 
\hat V | {\bf p} ~\mu, {\bf k-p+h} ~\mu 
\rangle $ of the Coulomb interaction 
which couples  the state $| {\bf k} ~\nu \rangle$, 
where the electron was initially excited by the incident  high-energy photon,
and the states $| {\bf h} ~\mu \rangle$, 
$| {\bf p} ~\mu \rangle$  and $| {\bf k-p-h} ~\mu \rangle$, which 
form the Mott band $\mu$. Now, we would like to estimate the transition 
rate $\Gamma_{\bf k}$ for an electron or hole of momentum ${\bf k}$ from 
a high-energy band to undergo the decay processes illustrated 
in Fig.~\ref{fig3}(b). Namely, Fig.~\ref{fig3}(b) illustrates the
following two processes:
(a) the initially photo-excited particle can decay into
two-quasiparticles plus one-quasihole  
given as 
$\alpha_{{\bf h}} \alpha^{\dagger}_{\bf p} \alpha^{\dagger}_{{\bf k - p - h}}
| 0 \rangle$ (first term in Fig.~\ref{fig3}(b)), or 
(b) the initially photo-created hole state  can 
decay into two-quasiholes plus one-quasiparticle 
given as $\alpha^{\dagger}_{{\bf h}} \alpha_{\bf p} 
\alpha_{{\bf k - p - h}}| 0 \rangle$ (second term in Fig.~\ref{fig3}(b)). 
For simplicity, the Mott-band index $\mu$ has been omitted as a label of
the $\alpha$'s.
Using Fermi's golden rule we can obtain the decay rate of the
bare electron (or hole) into two-quasiparticles plus one quasihole as:
\begin{eqnarray}
\Gamma_{\bf k}(\epsilon) = {{2\pi} \over {\hbar}} \sum_{{\bf p}, {\bf h}} 
|M|^2
\delta(\epsilon- E_{\bf k-p+h} - E_{\bf p} + E_{\bf h}),
\label{Fermi}
\end{eqnarray}
where $\epsilon\equiv E- 2 U^*$, and $M$ stands for the interaction matrix element 
which couples the initially excited particle
 state $c^{\dagger}_{{\bf k},\nu}| 0 \rangle$ 
($|0 \rangle$ is the ground state) to which the incoming solar photon 
couples directly, and the final two-quasiparticles and one-quasihole  given as 
$\alpha_{\bf h} \alpha^{\dagger}_{\bf p} \alpha^{\dagger}_{{\bf k - p - h}}
| 0 \rangle$, i.e.,
$\langle 0 | \alpha^{\dagger}_{\bf h}
\alpha_{{\bf p}} \alpha_{{\bf k-p+h}} \hat V c^{\dagger}_{{\bf k},\nu} 
| 0 \rangle$.  As can be inferred from  the main contribution to the
decay of the photo-excited high-energy electron-hole pair 
into two DO-hole pairs is due to the
two processes shown by the diagrams in Fig.~\ref{fig3}(b). 
Namely, when the Mott electron absorbs most of the photon
energy and it is excited to a high-energy band, it becomes possible 
for the electron to decay into two DOs and one hole in the Mott band.
When the created hole absorbs most of the photon energy, 
i.e., it corresponds to an excitation of an electron from an inner 
valence state to the Mott conduction band, it becomes possible for the
hole to decay into two holes from the Mott band and one DO
state from the Mott band. Each case can be approximated
by the calculation of the decay rate of the high-energy electron 
 or the decay rate of the  high-energy hole because the other
particle (the hole or electron respectively), in this order in 
perturbation theory, behaves only as a spectator. The later rates are
given by the above Eq.~\ref{Fermi}.
The matrix elements $M$ in this expression is given by
\begin{eqnarray}
M= { V \over N}  Z_{\bf h} Z_{\bf p} Z_{\bf k - p +h },
\label{Fermi2}
\end{eqnarray}
where the $Z's$ are the quasiparticle residues and $N$ is the
total number of lattice sites. 
The effect of the Hubbard interaction $U$ has been approximately taken into 
account by the $Z$, where $Z_{p} = \langle 0 | \alpha_p | {\bf p} ~\mu
\rangle$, and by using the renormalized Mott-band dispersion $E_{\bf k}$.
%becomes large the value of the matrix element is 
%significantly 
%smaller because of the presence of the product of the $Z's$ and $Z<1$.
In order to obtain an order of magnitude estimate for $\Gamma$,  we 
will take $V$ to be of the order of 1 eV.
In a Mott insulator, where the atomic limit is appropriate,
the charged highly-excited atom in the intermediate state of Fig.~\ref{fig3}(a)
is expected to interact rather strongly with the electrons of the other atoms
in its neighborhood.

This decay rate can be estimated by the following further approximation
\begin{eqnarray}
\Gamma_{\bf k}(\epsilon) \simeq {{2 \pi} \over {\hbar}} 
{V^2}{\bar Z}^6 D({\bf k},\epsilon),
\label{approximate}
\end{eqnarray}
where $\bar Z$ is the average quasiparticle/quasihole spectral
weight throughout the entire Mott-Hubbard band and
\begin{eqnarray}
D({\bf k},\epsilon)={1 \over {N^2}} \sum_{{\bf p}, {\bf h}} 
\delta(\epsilon- E_{\bf k-p-h} - E_{\bf p} + E_{\bf h}),
\end{eqnarray}
the two-particle/one-hole density of states. The approximation of 
Eq.~\ref{approximate}  is obtained by using the average $Z$ throughout the
integration.
\begin{figure}[htp]
\vskip 0.4 in
\includegraphics[width=3.275 in]{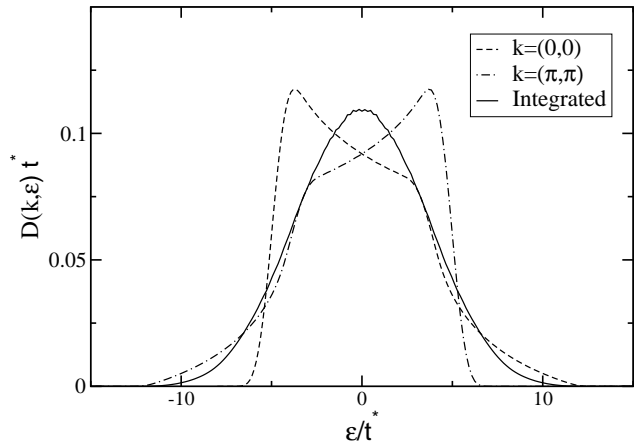}
\caption{The two-particle/one-hole density of states
for the simple tight-binding-like dispersion $E({\bf k})=-2t^*(\cos(k_xa)+
\cos(k_ya))$ for a square lattice.}
\label{fig4}
\end{figure}

The two-particle/one-hole density of states can be illustrated
using an effective tight-binding dispersion, i.e., 
$E_{\bf k}=-2t^* \sum_{\mu=1}^d \cos(k_{\mu}a_{\mu})$, where $d$ is the
number of dimensions and $k_{\mu}$, $a_{\mu}$, are the component of 
the momentum and the lattice constant along the direction $\mu$. 
In Fig.~\ref{fig3} we show $D({\bf k},\epsilon)$ for a square lattice.
Notice, that it is non-zero in an energy range of the order
of the single-particle/single-hole bandwidth $W=8 t^*$ and the value
of the density of states is of the order of $1/W$. The 
rate averaged over all momenta in the tight-binding band can be written as
\begin{eqnarray}
{\bar \Gamma}(\epsilon) &\simeq& {{2 \pi } \over {\hbar}} {{V^2}} 
{\bar Z}^6 \Delta(\epsilon),\label{approximate2}
\\
\Delta(\epsilon) &=&{{1} \over { N}} \sum_{\bf k} D({\bf k},\epsilon),
\end{eqnarray}
which is  the density of states 
given in Fig.~\ref{fig4} for two-dimensions (2D) averaged over
all momenta ${\bf k}$. Note that $\Delta(\epsilon)$ is of the order of 
$1/W$ in the energy 
range where there are available states.

The value of the residue $Z$ has been
estimated in the strong-coupling limit using the $t-J$ model 
in Ref.~\onlinecite{Liu} for 2D.
We have carried out the same calculation in 3D
and we found that the value of $Z$ is somewhat larger than in the 2D 
case, because the role of quantum fluctuations is less important in 3D.
Because we are interested in obtaining an approximate estimate of $\Gamma$,
we will use the results of Ref.~\onlinecite{Liu} for 2D.
%this estimate for $\Gamma$ will serve as a  lower bound
%for 3D because the actual value of $Z$ will be somewhat larger.  
Since the $t-J$ model is the
strong-coupling limit  of the Hubbard model  with  
$J/t=4t/U$, we need to choose $U \gg t$. Choosing $J/t=1$ (i.e., $U=4t$) 
from Table IV of
Ref.~\onlinecite{Liu} we obtain, $W =0.96 t$ and $Z=0.53$. 
Thus, taking $V=U=1$ eV yields $t=0.25$ eV, $W=0.24$ eV and 
$\Gamma \simeq 0.9 \times 10^{15} \sec^{-1}$ (using $\Delta(\epsilon) 
\sim 1/W$).
As can be inferred from the trend in Table IV of Ref.~\onlinecite{Liu}, 
larger values of $t/U$ will give larger values both for $Z$ and for the
bandwidth $W$ (similar to the value of the gap as needed) and 
consequently a larger $\Gamma$. However, while we believe this trend
to be qualitatively correct, the regime  $4t/U>1$ is not accessible
by the strong-coupling limit discussed in Ref.~\onlinecite{Liu}.
Nevertheless, the above estimate indicates that this process can be 
significantly faster than
the decay processes through phonon \cite{impact0,impact1,impact2,nozik} 
or spin-wave excitations which are expected to be of the order 
of $10^{12}-10^{13} \sec^{-1}$.
In addition,  the Mott-Hubbard gap and $W$ can be tuned to be 
several times smaller than the range of the energy of solar photons 
and, thus,  a photon has the energy required to
create more than one quasi-electron-hole pairs.

\section{Narrow-gap wide-band case}

\label{wideband} 

Now, let us consider the case of a narrow-gap but a wide-band
Mott-insulator, i.e.,
the case where $W \gg U^*$. 
It is believed
and supported by calculations that for any value of $W$
we can tune the value of $U$
to provide a small value of the gap $U^*$.
It is also known that the value of $U$ required to produce a small gap
should be greater than $W/2$.
Therefore, if we choose $W=8$ eV and $U^*=0.5$ eV we will need
a $U$  greater than $4$ eV. The exact value of $U$ is not required
for the following calculation, suffice it to know that it is 
greater than 4 eV.
When the
incident photon promotes an electron from the lower Hubbard-band
to the upper Hubbard band, the energy $E$ of the excited electron
can be large compared to the gap, say $E>nU^*$, and, thus, it can decay into
$n$ quasiparticles plus $n-1$ quasi-holes ($n=2,3,..$). 
The main process is depicted by the
diagram of Fig.~\ref{fig3}(b), where all 
the lines in this case denote quasiparticles/quasiholes of the same 
(Mott) band (i.e., $\nu=\mu$),
and the interaction is the Hubbard on-site Coulomb interaction U instead 
of the $V$ discussed previously.

In this case the
decay of the high-energy DO-hole pair,
produced by the solar photon, into two DO-hole pairs is due to the
two processes shown by the diagrams of
Fig.~\ref{fig3}(b) where the interaction is $U$ instead of $V$ and
the decay rate is given by Eq.~\ref{Fermi} with the matrix
element $M$ given by Eq.~\ref{Fermi2} where $V$ is replaced by 
the Hubbard on-site Coulomb repulsion $U$. 
Namely, (b) when the electron absorbs most of the photon
energy and it is excited to the upper-Hubbard band to become 
a high-energy DO site, it becomes possible 
for the DO to decay into two DO states and one hole.
(b) When the created hole absorbs most of the photon energy, 
 it becomes possible for the
hole to decay into two holes in the Mott band and one DO
state in the Mott band. Here also, either (a) or (b) can be
estimated 
by the calculation of the decay rate of the high-energy DO 
 or the decay rate of the  high-energy hole because the other
particle (the hole or DO respectively), behaves only as a spectator. 

Therefore, an estimate of the decay rate
is given by Eq.~\ref{Fermi} by replacing $V$ with $U$ which can
be approximated by $\hbar \Gamma \sim 2\pi U^2 {\bar Z}^6/W$ (See discussion
leading to Eq.~\ref{approximate2} starting from Eq.~\ref{Fermi}).
Using this expression and $W=8$ eV and $U=4$ eV,  the lower bound
required to yield $U^*=0.5$ eV, we also obtain a large  
$\Gamma \sim 0.5 \times 10^{15} sec^{-1}$
(taking the same value of $\bar Z$ used in the previous paragraph). 

\section{Issues regarding efficiency}

\subsection{Impact ionization, Auger recombination and
other processes}
Impact ionization increases the number of charge carriers per absorbed
phonon. However, Auger recombination which is the
inverse process should not be omitted.
First, let us begin our discussion from the open-circuit situation.
We found that for the type of Mott insulators discussed
in the previous section, the characteristic time
scale for thermalization of the photo-excited electrons
and holes is much shorter than the time needed for 
phonons to thermalize the electronic system with the lattice.
Therefore, it is reasonable to assume 
that the photo-created DOs and the holes in a Mott insulator
thermalize among themselves at a temperature $T_e$ higher than
the lattice temperature $T_L$, by means of the electron-electron
interaction (the interaction $V$ defined in the previous section
in the case of narrow-band or the Hubbard on-site $U$ in the case of 
a wide-band Mott-insulator). This interaction between the
electronic degrees of freedom leads to processes such as
impact ionization and Auger recombination, as well as
carrier-carrier scattering. This equilibrium state is achieved
as follows. 

The energy current of the absorbed solar photons
is given by
\begin{eqnarray}
J_{absorbed} = {{\Omega_s} \over { 4 \pi^3 \hbar^3 c^2} }
\int_{U^*}^{\infty}
d\epsilon {{\epsilon^3} \over {\exp({{\epsilon} \over {k_B T}}) -1}}
\end{eqnarray}
where  $\Omega_s$ is the solid angle under which the Sun is seen, 
and $T$ is the
Sun's surface temperature $T\simeq 5760$ K. Here, the lower energy cutoff
is set by the value of the Mott gap $U^{*}$.

First, there is the impact ionization process, and, then,
the Auger recombination process, i.e., the inverse
of the impact ionization process, shown as the reverse process
of the one shown in Fig.~\ref{fig3}. These two processes can be also
denoted as follows 
\begin{eqnarray}
{\rm {DO}}_1 \leftrightarrow {\rm {DO}}^{\prime}_1 + {\rm {DO}}_2 + {\rm H},
\label{impact1}
\end{eqnarray}
where DO$_1$ is the initial DO which creates the additional DO-H pair in 
the Mott-insulator denoted as
DO$_2$, H, by changing its energy-momentum state to become DO$^{\prime}_1$.
In addition, there is the analogous processes for high energy holes, i.e.,
\begin{eqnarray}
{\rm {H}}_1 \leftrightarrow {\rm {H}}^{\prime}_1 + {\rm {DO}} + {\rm H}_2.
\label{impact2}
\end{eqnarray}
Furthermore, there is the
electron-electron scattering (due to the  matrix element $V$ defined in 
Section~\ref{narrowband} or the
on-site Coulomb repulsion $U$), 
which leads to DO-DO, hole-hole, and DO-hole interaction processes:
\begin{eqnarray}
{\rm {DO}}_1 + {\rm {DO}}_2 &\leftrightarrow&  {\rm {DO}}^{\prime}_1
+ {\rm {DO}}^{\prime}_2, \label{electron-electron1}
\\
{\rm {H}}_1 + {\rm {H}}_2 &\leftrightarrow&  {\rm {H}}^{\prime}_1
+ {\rm {H}}^{\prime}_2, \label{electron-electron2}
\\
{\rm {DO}} + {\rm {H}} &\leftrightarrow&  {\rm {DO}}^{\prime}
+ {\rm {H}}^{\prime}.
\label{electron-electron3}
\end{eqnarray}
If these scattering processes are excluded, and we consider only
impact ionization and Auger recombination, 
then, only those processes involving DOs and holes in certain energy
and momentum range which allow for impact ionization and its
inverse through energy-momentum conservation would take place
and everything else would be excluded. As a result, the final 
distribution of carriers would depend on energy and momentum and, thus,
it cannot
be described by a Fermi-Dirac distribution. Therefore, in the absence
of electron-phonon interaction, the above
scattering processes are important in order to establish a common
temperature and a Fermi-Dirac distribution.
 
If we now consider the equilibrium of all of the 
above processes, we realize that  they produce 
a distribution of DO-H pairs in equilibrium with a distribution of
photons over energy due to the emission process:
\begin{eqnarray}
DO + H \leftrightarrow \gamma.
\label{emission}
\end{eqnarray}
in which a DO and a hole (H) combine and a photon ($\gamma$) is emitted 
(luminescence) or the inverse where a photon produces a DO-H pair
in the Mott insulator.
This distribution of photons in equilibrium with 
a gas of quasiparticles consisted of DOs and holes 
is analogous to the case of electrons and holes in 
a band-insulator which can be described \cite{Wurfel1,Wurfel} by a thermal
distribution of photons at a temperature $T_e$ and
a chemical potential $\mu_{\gamma}=\mu_{DO-H}=0$ 
(see Refs.~\onlinecite{Wurfel1,Wurfel}). The value
of $T_e$ is determined from the equation
\begin{eqnarray}
J_{emitted} &=& J_{absorbed},\label{equilibrium}
 \\
J_{emitted} &=& {{\Omega_e} \over { 4 \pi^3 \hbar^3 c^2} }
\int_{U^*}^{\infty}
d\epsilon {{\epsilon^3} \over {\exp({{\epsilon-\mu_{\gamma}} \over 
{k_B T_e}}) -1}}.
\end{eqnarray}
Here $\Omega_e$ is the solid angle of the emitter 
and it is $\pi$ for a planar emitter.
This equation defines a temperature $T_e$ of the electronic
system which is considered decoupled from the lattice and the phonons.
A value of $\mu_{DO-H}=0$ is expected for a system in which the particle
number is not conserved as is the case for the DO and holes in the 
equilibrium state of impact ionization and Auger-recombination.

Assuming that the density of the DO and of holes near the interface
is not large, the above process produces an equilibrium 
Fermi-distribution of DO and holes at the above temperature
$T_{e}$ with no separation of the quasi-Fermi-energies, i.e, the chemical
potential difference between DO and holes $\mu_{DO-H}$ is zero.

Now, equilibrium is achieved
(under open circuit conditions) at the
temperature $T_e$ which is higher than the lattice temperature
$T_L$. As shown in the previous sections, the time scale 
for the Mott insulator to reach equilibrium (through impact ionization,
 Auger recombination and electron-electron scattering) is much 
shorter that the time scale required for such a system to reach thermal
equilibrium with the lattice through the process
\begin{eqnarray}
DO + H \leftrightarrow \Gamma
\label{phonon}
\end{eqnarray}
where $\Gamma$ denotes a phonon. This is so because of the facts
that i) the time scale for the decay processes described by 
Eqs.~\ref{impact1} and ~\ref{impact2} is much faster than the process
of electron decay via phonons and ii) the interaction matrix elements
involved in  the processes described by Eq.~\ref{electron-electron1},
\ref{electron-electron2},\ref{electron-electron3}
are much stronger than the electron-phonon interaction matrix
elements leading to the processes \ref{phonon}.
These facts allow first equilibration of the electronic system 
to a temperature $T_e$, which is different than $T_L$ by creating
multiple DO-H carriers per incident photon. This has been argued
in Refs.~\onlinecite{Werner,Brendel,Spirkl} to lead to a significant increase of
the efficiency by means of carrier multiplication \cite{Brendel,Spirkl}
for a conventional solar cell.
By treating the carriers in the Mott-insulator as weakly
interacting quasiparticles which obey Fermi-Dirac statistics, which may allow
us to apply the same assumptions and approximations used by Brendel 
et al.\cite{Brendel} to calculate the efficiency of a conventional solar
cell, we also find that the maximum efficiency is bounded from above by 
the maximum of the following function of the voltage $V$ and the Mott gap $U^*$
\begin{eqnarray}
\eta(V,U^*) &=& q V {{ g(U^*)- \xi ~  r(V,U^*)} \over {P_{in}}}, \\
g(U^*) &=& \int_{U^*}^{\infty} d\epsilon  
{{m(\epsilon) \epsilon^2} \over {\exp(\epsilon/k_BT_e)-1}}, \\
r(V,U^*) &=& \int_{U^*}^{\infty} d\epsilon 
{{ m(\epsilon) \epsilon^2} \over {\exp((\epsilon-q V)/k_BT_e)-1}}, \\
P_{in} &=& \int_{0}^{\infty} d\epsilon 
{{\epsilon^3} \over {\exp(\epsilon/k_BT_s)-1}}.
\end{eqnarray}
In the above expression, apart from a multiplicative constant 
common in the numerator and denominator expressions which cancels out,  
$g(V)$ corresponds to the generation current, $r(V,U^*)$ to the recombination
current and $P_{in}$ the total power carried by the sunlight; 
$m(\epsilon)=min([\epsilon/U^*],m_0)$ is the number of DO-H pairs 
corresponding to each photon in dynamic equilibrium, 
where $[\epsilon/U^*]$ stands for the integer part
of  $\epsilon/U^*$
and $m_0$ is a maximum  allowed value of the carrier multiplication. 
Furthermore, $q$ is the carrier charge and
$\xi=\pi/\Omega_s$ is the ratio of $\pi$ to the solid
angle through which the Sun is seen from the Earth and the case $\xi=1$ 
corresponds to fully concentrated sunlight.
If we follow Werner et al.\cite{Werner} and Brendel et al.\cite{Brendel} 
and use a value of $T_e=300 K$,
we reproduce the same upper bounds for the solar-cell efficiencies 
as a function of $U^*$ as those reported 
in Fig.~1 and Fig.~2 of Ref.~\onlinecite{Brendel}.
The calculated values for these upper bounds for solar-cell efficiencies
reach up to 85 \% for fully concentrated sunlight and in the
limit of vanishing value of $U^*$ when $m_0$ becomes very large. 

\subsection{Further efficiency improvement}

As argued by W\"urfel \cite{Wurfel}, however,
since the phonons are excluded from the equilibration process,
the DO and holes through Eq.~\ref{equilibrium} will reach an
equilibrium temperature $T_e$ which should be higher than 300 K;
as a result of the fact that the equilibrium temperature of the electronic
degrees of freedom is higher than 300 K assumed in the calculation
of Brendel el al.\cite{Brendel}, the efficiency should be lower than the
values calculated using $T_e=300$ K.
As was argued in the previous sections, the electronic equilibration is 
achieved much faster
than the equilibration with the lattice, and this is important 
for another reason, namely, it allows the application of an
idea discussed by W\"urfel \cite{Wurfel}, which is summarized below,
in order to improve the efficiency further.

The photovoltaic cell can achieve maximum efficiency
when there is a way to cool the gas of carriers to the lattice
temperature $T_L$ via an isentropic process \cite{Wurfel}
where the relatively high energy of the carriers is not dissipated
by heating the lattice but rather by creating
electrical energy. Following W\"urfel \cite{Wurfel} we assume that
(a) impact ionization (and Auger recombination) occurs at much higher
rate than photon absorption/emission, which as argued above is more likely
in a Mott-insulator based solar-cell and (b) the carriers are removed
from the solar-cell through membrane leads which are attached to
the solar-cell on each side \cite{Wurfel} and which can be
semiconductors of n-type and p-type
 with narrow bands of band-width $\delta \epsilon$ 
and large band-gaps.

Because of the assumption (a) discussed above, the equilibrium 
of the carriers is not disturbed
by the withdrawal of electrons and holes when the circuit is short.
Now, the interaction with the slower acting phonons of the membranes
cannot be avoided; in this case, however, this interaction is 
utilized in a favorable way. Namely, the hot electrons and holes 
with energies $\epsilon_e$ and $\epsilon_h$  in a narrow interval 
$\delta \epsilon$ and with temperature $T_e$ 
can be cooled to the temperature of the lead-membranes $T_L$ by keeping the
entropy constant.
This can occur due to the above mentioned assumption (b) where the carriers, 
whose concentration is constant,
are transported via an approximately flat band, which prevents
energy losses because the carriers keep approximately the same
energy. This works as a Carnot engine which produces work
$\mu_{eh}=(1-T_L/T_e) (e_e+e_h)$ which appears
as a difference of the quasi-Fermi levels between the electron and hole
carriers in the two membranes and is given by $q V$.
The details of how this works
as well as the calculation of the efficiency can be
found in Ref.~\onlinecite{Wurfel}. This calculation leads 
to a maximum energy efficiency of 85 \%  also in the limit of a 
vanishing Mott-gap 
and for maximum concentration of solar cell radiation.
Therefore, in this subsection, we conclude that the idea proposed by W\"urfel
may be utilized in solar-cells based on Mott-insulators.

\subsection{Other issues regarding efficiency}

A significant reason why a narrow-band Mott-insulator
may be preferred over the wide-band case is that
the threshold for impact ionization to occur for the latter
case (and in a wide-band semiconductor\cite{Shockley,Vavilov,Tauc})
 is approximately 3 times the Mott-insulator gap.
This is because  energy and momentum are conserved together
via the impact ionization (or Auger recombination) and presence of
a wide-band implies a strong dependence of the energy with momentum. 
On the other hand, a narrow-band Mott insulator of the case
discussed in Sec.~\ref{narrowband} gives a much smaller
energy threshold for impact ionization.

Finally, there is an important issue to address which is that 
in a Mott insulator the carrier mobility
might be low. However, the effective 
collection length, as well as the drift field of the minority carriers
can be large  in such a Mott insulator, due to the
following reasons.  The main reason why a DO 
diffuses from the n-doped region inside the p-region to fill a hole, is 
the large energy gain which will be resulted by the fact that 
the double occupancy in a Mott insulator of the type considered here, 
costs several eV. These materials become Mott insulators precisely 
because of this large on-site Coulomb repulsion which prevents the 
electrons from hopping from one site to a nearest neighboring site and this
leads to an insulator in the half-filled band case.
Therefore, this strong on-site Coulomb repulsion between electrons
produces high diffusivity because the DO in the n-doped region 
move to the p-doped region to gain a large amount of energy by filling the
hole and this may lead to a high drift voltage and large diffusion length
regardless of the fact that the electron mobility might be low.

\section{Conclusions}

What should be the nature of the absorption spectrum in the case
of the Mott insulator considered here in order for it to be a candidate for
the basis of an efficient solar cell? 
There are two cases as discussed in the
paper for the case of a narrow gap Mott insulator. 
a) Narrow Mott-band materials. In this case in order for the
effect considered in the present paper to be significant, we need
other bands present, not necessarily of Mott type.
The incident solar photon may be absorbed by an electron from
another such valence band or the Mott conduction band and be
promoted to a highly-excited band.  If there are no issues related
to selection rules, these electrons are expected to decay very quickly 
(for the reasons discussed in the paper) to produce one or multiple 
DO-hole pairs from the Mott band.
Therefore, if the material is appropriately selected,
a large fraction of the solar spectrum, could be absorbed
leading to a much larger photo-current than in 
band-semiconductors. b) In a wide-band Mott insulator,
almost the entire solar spectrum could be absorbed
by the same single Mott-band. As analyzed in the present paper,
in this case the initially photo-excited single DO-hole
pair decays into two or more DO-hole pairs by means of the 
on-site Coulomb repulsion within a time-scale much shorter
than other internal processes which dissipate the excess
energy of the DO or hole beyond the Mott gap.

In conclusion, it was argued that 
a p-n heterojunction in which the illuminated side is a
narrow-gap Mott insulator,
if fabricated, should be expected to behave as a high quantum 
efficiency solar cell. This can be achieved when the 
Mott insulator's energy spectrum
for multiple quasi-electron-hole pairs falls within the energy range 
of solar photons. We find that in such case, 
a relatively high-energy incident
photon initially creates a high-energy electron-hole pair which 
quickly dissociates into multiple quasi-electron-hole pairs  
and this leads to an increased 
photo-current. As required for
the device to work, it is estimated that this dissociation occurs in a 
time scale shorter
than the time needed for the excited electron-hole pair to relax
through the emission of phonons or other intrinsic excitations.
This allows equilibration of the electronic degrees of freedom
among themselves, with excitation of multiple DO-H
pairs for every absorbed photon, 
at a different temperature, on a time
scale significantly shorter than the time required for the
system to reach equilibrium with the lattice.
This separation of time scales also allows the application of other 
previously proposed ideas to improve solar-cell efficiency by cooling  the 
carriers isentropically to a lower temperature\cite{Wurfel}.

\section{Acknowledgments}
I wish to thank V. Dobrosavljevic, B. I. Halperin, 
C. S. Hellberg, P. Schlottmann, and
S. von Moln\'ar for useful discussions.

\end{document}